\documentclass[AMA,STIX2COL,Linenumbersoff]{MRM}
\articletype{Article Type}%

\received{}
\revised{}
\accepted{}
\begin{document}

\title{Quantitative Macromolecular Proton Fraction Imaging using Pulsed Spin-Lock\protect\thanks{Qianxue Shan and Ziqiang Yu contributed equally to this work.}}

\author[1]{Qianxue Shan}{\orcid{0000-0003-1975-6011}}

\author[1]{Ziqiang Yu}{\orcid{0009-0000-2466-4942}}

\author[1,2]{Baiyan Jiang}{\orcid{0000-0002-2732-2152}}

\author[1]{Jian Hou}{\orcid{0000-0003-1746-4138}}

\author[1]{Qiuyi Shen}{\orcid{0009-0005-4337-3813}}

\author[1]{Winnie Chiu-Wing Chu}{\orcid{0000-0003-4962-4132}}

\author[3]{Vincent Wai-Sun Wong}{\orcid{0000-0003-2215-9410}}

\author[1]{Weitian Chen}{\orcid{0000-0001-7242-9285}}

\authormark{Qianxue Shan and Ziqiang Yu \textsc{et al}}

\address[1]{\orgdiv{Department of Imaging and Interventional Radiology}, \orgname{the Chinese University of Hong Kong}, \orgaddress{\state{Hong Kong SAR}, \country{China}}}

\address[2]{\orgname{Illumination Medical Technology Limited}, \orgaddress{\state{Hong Kong}, \country{China}}}

\address[3]{\orgdiv{Department of Medicine \& Therapeutics}, \orgname{the Chinese University of Hong Kong}, \orgaddress{\state{Hong Kong SAR}, \country{China}}}

\corres{Weitian Chen, Department of Imaging and Interventional Radiology, the Chinese University of Hong Kong, Room 15, Sir Yue Kong Pao Centre for Cancer, Prince
of Wales Hospital, Shatin, NT, Hong Kong SAR, China. \email{wtchen@cuhk.edu.hk}}

\finfo{ \fundingAgency{Research Grants Council of the Hong Kong SAR: Hong Kong General Research Fund (GRF)} Grant/Award Number\fundingNumber{14201721}.
\fundingAgency{Innovation and Tchnology Commission of the Hong Kong SAR} Grant/Award Number \fundingNumber{MRP/046/20x}.}

\abstract[Abstract]{\section{Purpose} 
Recent studies have shown that spin-lock MRI can simplify quantitative magnetization transfer (MT) by eliminating its dependency on water pool parameters, removing the need for a T1 map in macromolecular proton fraction (MPF) quantification. However, its application is often limited by the requirement for long radiofrequency (RF) pulse durations, which are constrained by RF hardware capabilities despite remaining within specific absorption rate (SAR) safety limits.
\section{Methods}
To address this challenge, we propose a novel method, MPF mapping using pulsed spin-lock (MPF-PSL). MPF-PSL employs a pulsed spin-lock train with intermittent free precession periods, enabling extended total spin-lock durations without exceeding hardware and specific absorption rate limits. A comprehensive analytical framework was developed to model the magnetization dynamics of the two-pool MT system under pulsed spin-lock, demonstrating that MPF-PSL achieves MT-specific quantification while minimizing confounding effects from the water pool.
The proposed method is validated with Bloch-McConnell simulations, phantoms, and in vivo studies at 3T.
\section{Results}
Both Bloch-McConnell simulations and phantom validation demonstrated that MPF-PSL exhibits robust insensitivity to water pool parameters while enabling high-SNR MPF quantification. In vivo validation studies confirmed the method's clinical utility in detecting collagen deposition in patients with liver fibrosis.
\section{Conclusion}
MPF-PSL presents a practical solution for quantitative MT imaging, with strong potential for clinical applications.
}

\jnlcitation{\cname{%
\author{Qiaxue Shan}, 
\author{Ziqiang Yu}, 
\author{Baiyan Jiang}, et al.} 
(\cyear{2025}), 
\ctitle{Quantitative Macromolecular Proton Fraction Imaging using Pulsed Spin-Lock}, \cjournal{Magn. Reson. Med.}, \cvol{2025;00:1--6}.}

\maketitle

\section{Introduction}\label{secIntro}

Magnetization transfer (MT) imaging provides macromolecular-related contrast by detecting the exchange of magnetization between free water protons and semisolid macromolecular protons\cite{wolff1989mtc, henkelman2001mt}. This noninvasive technique offers unique sensitivity to microscopic tissue properties, enabling the characterization of molecular changes that are often undetectable with conventional imaging methods. MT imaging has been extensively investigated in neuroimaging, where it shows a strong correlation with myelin histopathology and disease severity, making it a valuable tool for studying demyelinating diseases such as multiple sclerosis\cite{schmierer2004ms, yarnykh2015ms, filippi2007ms}. Beyond neuroimaging, MT imaging has shown significant promise in other clinical applications. For example, recent studies have underscored its potential in the assessment of liver fibrosis\cite{yarnykh2015liver, hou2023detecting, wilczynski2023MEX}, offering new opportunities for the noninvasive evaluation of chronic organ diseases.

Various methods have been developed to quantify MT effects. The MT ratio (MTR) \cite{wolff1989mtc,dousset1992mtr} is a simple and widely used approach that measures MT effects by comparing signal intensities acquired before and after saturation. However, MTR is highly dependent on experimental conditions and pulse sequence parameters, which limits its ability to reliably reflect intrinsic tissue properties.
Quantitative magnetization transfer (qMT) methods address these limitations by estimating tissue-specific parameters, such as the macromolecular proton fraction (MPF), relaxation rates, and exchange rate constants, using a two-pool model \cite{henkelman1993poolmodel,morrison1995poolmodel}. A commonly used class of qMT methods relies on off-resonance saturation\cite{henkelman1993poolmodel,yarnykh2002pulsed,sled2001qMT,yarnykh2012fast}, where tissue parameter maps are derived by analyzing saturation-induced signal reductions in the water pool and fitting the data to the two-pool model. Over the past few decades, a variety of alternative qMT techniques have been developed\cite{gochberg2007InversionRecovery,dortch2018optimization,gloor2008quantitative,hilbert2020fingerprinting,kang2022fingerprinting,hou2020mpfsl}. However, despite the ability to quantify tissue-specific parameters, qMT methods are rarely implemented in routine clinical imaging due to challenges in data acquisition and post-processing.

The two-pool model commonly used in qMT imaging to quantify tissue parameters often couples the contributions of the water pool and the macromolecular pool within the quantification model. This coupling increases the complexity of both data acquisition and post-processing. For example, conventional saturation-based qMT methods require the acquisition of an additional $T_\mathrm{1}$ map or rely on specific assumptions about tissue parameters. Recently, qMT based on the spin-lock technique (hereafter referred to as qMT-SL) has been proposed as a solution to these challenges \cite{hou2020mpfsl}. By decoupling the water pool and the MT pool in the quantification model, the qMT-SL approach enables a model that is specific to the MT pool. This method has been successfully demonstrated for MPF mapping\cite{hou2020mpfsl,hou2023detecting,hou2024brain}. In addition, recent studies have reported that qMT quantification can be confounded by the orientation of tissue structures\cite{pampel2015orientation,karan2024orientation}. The qMT-SL approach offers the potential to address this limitation. By utilizing an off-resonance spin-lock, it achieves sufficiently high spin-locking fields to suppress residual dipolar coupling without introducing direct water saturation or violating specific absorption rate (SAR) limit, thereby enabling orientation-independent MPF quantification in ordered tissue structures \cite{gao2024orientation}.

{\color{black}For qMT-SL, it is critical to use a relatively long spin-lock duration to achieve reliable quantification. A major barrier to clinical implementation is the limited spin-lock duration achievable under typical $B_\mathrm{1}$ amplitudes (i.e., >300 Hz) due to hardware constraints, which may cause scan failures even when SAR remains within safety limits. RF power deposition increases with the square of the magnetic field strength and RF coil size\cite{smith2006pulsed}. When RF pulses exceed the limits of the RF power amplifier (RFPA), the RFPA may deplete during operation, leading to sequence failure.
Commercial MRI systems are often not optimized for continuous-wave RF pulses with relatively large $B_\mathrm{1}$ amplitudes, making continuous-wave spin-lock experiments particularly challenging. Abdominal scans using body coils further increase RF hardware demands compared to head or extremity scans, as the larger body size requires higher RF power\cite{smith2006pulsed}. While lower field strengths reduce SAR concerns, they often face greater hardware limitations because RF amplifiers are designed to match system power budgets. For instance, 1.5 T systems may face more RF power constraints than 3 T systems \cite{chan2019pulse}, and low-field systems (e.g., 0.55 T) experience even more severe restrictions.
To address these challenges, we propose a pulsed spin-lock approach that utilizes spin-lock modules interspersed with pauses, allowing the RF amplifier to recharge between spin-lock RF pulses.}

Although the solution for the MT dynamics of qMT-SL under CW spin-lock has been reported in previous studies\cite{hou2020mpfsl, hou2024brain}, these solutions are not applicable to qMT-SL using pulsed spin-lock RF trains. {\color{black}On the other hand,} Gochberg \emph{et al.} introduced an analytic solution for spin dynamics in pulsed CEST imaging \cite{gochberg2018AnalyticPulsed,lankford2022CESTana}, based on the chemical exchange rotation transfer (CERT) framework\cite{zu2014cert}. Roeloffs \emph{et al.} proposed a solution for CEST quantification under pulsed spin-lock acquisitions \cite{roeloffs2015ISAR}. 
{\color{black} However, these methods are specific to CEST imaging and cannot be directly applied to qMT-SL due to fundamental differences in their underlying physics and mathematical models. Specifically, MT occurs over a wide range of frequency offsets, and the exchange involving transverse magnetization components can be neglected due to the extremely short $T_\mathrm{2}$ of the macromolecular pool. In contrast, CEST occurs only at specific frequency offsets, and both transverse and longitudinal magnetization are considered during the exchange due to the relatively long $T_\mathrm{2}$ of the CEST pool. While the CEST model uses a 6×6 matrix with Lorentzian lineshape, the MT model employs a 4×4 matrix and supports non-Lorentzian lineshapes, such as Gaussian or super-Lorentzian, to better describe experimental data\cite{morrison1995poolmodel}.}

A solution for the two-pool model in qMT-SL using pulsed spin-lock is essential to fully exploit this technology, yet it presents significant challenges. {\color{black} In this work, we introduce a transient method for implementing qMT-SL using off-resonance pulsed spin-lock, supported by a comprehensive theoretical framework that includes derived analytical solutions for the spin dynamics of the two-pool MT model. This technique provides a relaxation rate, $R_\mathrm{mpfsl,pul}$, that does not depend on water pool parameters, allowing for the determination of the MPF without the need for a $T_\mathrm{1}$ map.} Our experiments demonstrate that, compared to the conventional CW approach, the proposed method significantly improves the {\color{black}reliability, consistency, and accuracy of the measurements} by enabling longer total spin-lock durations under typical RF hardware and SAR constraints. This advancement greatly enhances the feasibility and clinical applicability of qMT-SL methods.

\begin{figure*}[t]
    \centering\includegraphics[width=1\linewidth]{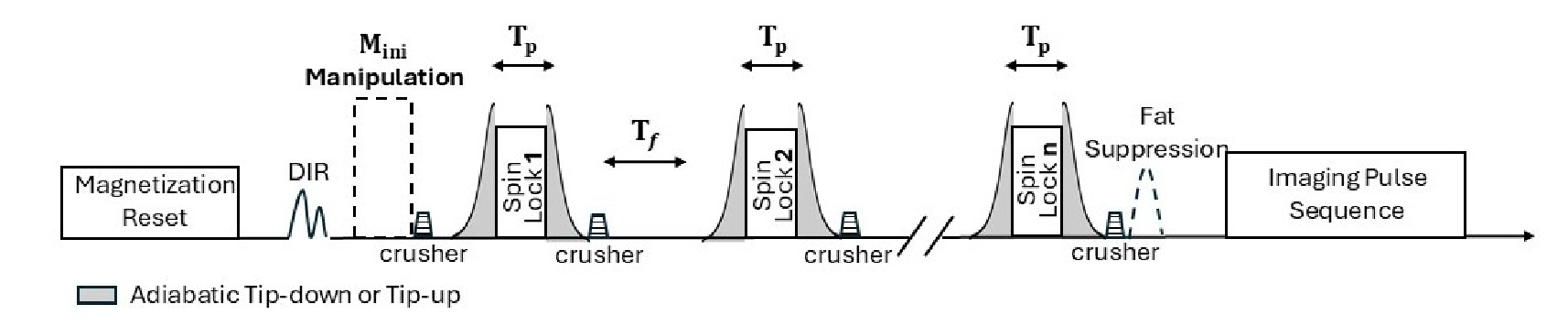}
    \caption{{\color{black} The schematic diagram of the proposed pulsed spin-lock sequence. The sequence consists of n spin-lock RF pulses, each with a duration of $T_\mathrm{p}$, separated by n-1 free precession periods, each lasting $T_\mathrm{f}$. The magnetization is first tipped into alignment with the spin-lock direction using an adiabatic RF pulse, and after the spin-lock, it is returned to the z-direction with another adiabatic RF pulse. Crusher gradients are applied following each spin-lock RF pulse to de-phase the transverse magnetizations. Other magnetization settings resemble those in original MPF-SL methods.}}
    \label{fig:sequence}
\end{figure*}

\section{Theory}
{\color{black} The pulse sequence for the proposed pulsed spin-lock approach is detailed in Fig. \ref{fig:sequence}. It is  composed of repeated spin-lock modules, where each module includes two stages: the CW irradiation stage, lasting $T_\mathrm{p}$ and the free precession stage, lasting $T_\mathrm{f}$.} The magnetization dynamics of the pulsed spin-lock sequence are derived as follows.

During the CW irradiation stage, we adopt a monoexponential relaxation model governed by the longitudinal relaxation rate in the rotating frame, $R_\mathrm{1\rho}$, as validated by previous studies\cite{zaiss2013eigenvalue,zaiss2015combined}. For the free precession stage, a biexponential decay model is used to account for the magnetization transfer in the absence of RF irradiation. This model incorporates the rotational exchange effects \cite{zu2014cert, gochberg2018AnalyticPulsed} between longitudinal magnetization of two pools. A transient state relationship between the longitudinal magnetizations of the two pools is derived for the CW irradiation stage and is used to link the two stages within a single spin-lock module. The magnetization dynamics of an individual spin-lock module, encompassing both stages, are then determined using a geometric series approach. This framework enables modeling of the overall magnetization behavior and calculation of spin dynamics throughout the entire pulsed spin-lock sequence.

\subsection{Magnetization Dynamics Under CW Spin-Lock}

The magnetization dynamics in the presence of magnetization transfer can be modeled using the Bloch-McConnell (BM) equations, which account for magnetization transfer effects between the free water pool (pool A) and the semi-solid macromolecular pool (pool B).
Each pool is characterized by its own magnetization vector, longitudinal relaxation rate $R_\mathrm{1,a/b}=1/T_\mathrm{1,a/b}$ and transverse relaxation rate $R_\mathrm{2,a/b}=1/T_\mathrm{2,a/b}$, and exchange rates between pools $k_\mathrm{ab}=f_\mathrm{b}k_\mathrm{ba}$, where $f_\mathrm{b}$ is the pool size fraction of pool B. The macromolecular proton fraction is defined as:
\begin{equation}
    \mathrm{MPF}=\frac{f_\mathrm{b}}{(1+f_\mathrm{b})}.
\end{equation}
The macromolecular pool's MT effects can be incorporated into the BM equations via the saturation rate term $R_\mathrm{rfb}$ term, which accommodates distinct lineshapes including Gaussian functions for solids and gels, and super-Lorentzian functions for living tissues\cite{zaiss2015combined,hou2020mpfsl}.

Using the analytical solution of BM equations,
the resulting magnetization is found to exhibit a mono-exponential decay aligned with the effective field, tiled by an angle $\theta$, and dominated by the rate $R_\mathrm{1\rho}$ in the rotating frame,
\begin{equation}
    M_\mathrm{za}(t)=M_\mathrm{za}(0)\mathrm{e}^{-R_\mathrm{1{\rho}}t}+ M_\mathrm{za}^\mathrm{ss} (1-\mathrm{e}^{-R_\mathrm{1\rho}t}),
\label{eqcw}
\end{equation}
where $M_\mathrm{za}^\mathrm{ss} = M_\mathrm{0a}R_\mathrm{1a}\cos\theta /{R_\mathrm{1\rho}}$ is the steady-state magnetization with $\theta = \text{atan} \left(\omega_1 /\Delta\omega\right)$. $\omega_\mathrm{1}$ is the amplitude of the RF pulse and $\Delta\omega$ is the resonance frequency offset. $M_\mathrm{za}(0)$ denotes the initial magnetization.

In this two-pool model, the longitudinal relaxation under CW spin-lock in the rotating frame $R_\mathrm{1\rho}$ can be expressed as a superposition\cite{zaiss2015combined}:
\begin{equation}
    R_\mathrm{1\rho}(\Delta\omega,\omega_1)=R_\mathrm{water}(\Delta\omega,\omega_1)+R_\mathrm{mt}(\Delta\omega,\omega_1).
    \label{eqR1rho}
\end{equation}
The terms $R_\mathrm{water}$ and $R_\mathrm{mt}$ refer to the effective relaxation rate of the free water pool and the MT-dependent relaxation rate in the rotating frame, respectively. The $R_\mathrm{water}$ can be calculated using the following equation,
\begin{equation}
    R_\mathrm{water}(\Delta\omega,\omega_1) = R_\mathrm{1a}\cos^2\theta+R_\mathrm{2a}\sin^2\theta.
    \label{eqRwater}
\end{equation}
{\color{black}The detailed expression for $R_\mathrm{mt}$ is provided in Supporting Information: Derivation S1, based on the eigenvalue approximation described by Zaiss \emph{et al.}\cite{zaiss2015combined}.}

\subsection{Magnetization Dynamics During Free Precession}
In the free precession stage, where RF irradiation is absent, the magnetization evolution remains described by 
BM equations, but with $\omega_1=0$ and $\Delta\omega=0$.
In this case, the z-component decouples from the transverse components, leading to a simplified bi-exponential model for the 
longitudinal magnetization.

Previous studies \cite{edzes1977cross, gochberg1998InversionRecovery, gochberg1999InversionRecovery} have solved the bi-exponential dynamics of the free BM equations, which were later simplified by Gochberg \cite{gochberg2018AnalyticPulsed} through a first-order Taylor series approximation in $f_\mathrm{b}$. The resulting magnetization evolution, $M_\mathrm{za}(t)$, is expressed as:
\begin{equation}
\begin{split}
M_{\mathrm{za}}(t) & =  \left[f_\mathrm{b}\mathrm{e}^{-k_\mathrm{ba}t} + (1-f_\mathrm{b})\mathrm{e}^{-R_\mathrm{1a}t}\right]M_\mathrm{za}(0) \\ &+ \left[-f_{\mathrm{b}}\mathrm{e}^{-k_\mathrm{ba}t} + f_{\mathrm{b}}\mathrm{e}^{-R_\mathrm{1a}t}\right] M_\mathrm{zb}(0)M_\mathrm{0a}/M_\mathrm{0b} \\ 
&+ (1 - \mathrm{e}^{-R_\mathrm{1a}t})M_\mathrm{0a},
\end{split}
\label{eqbiexp}
\end{equation}
which is the magnetization after a bi-exponentional recovery at time $t$ and initial magnetization $M_\mathrm{za}(0)$ and $M_\mathrm{zb}(0)$.

\subsection{Magnetization Dynamics During combined CW Spin-Lock and Free Precession}
\begin{figure}
    \centering
    \includegraphics[width=1\linewidth]{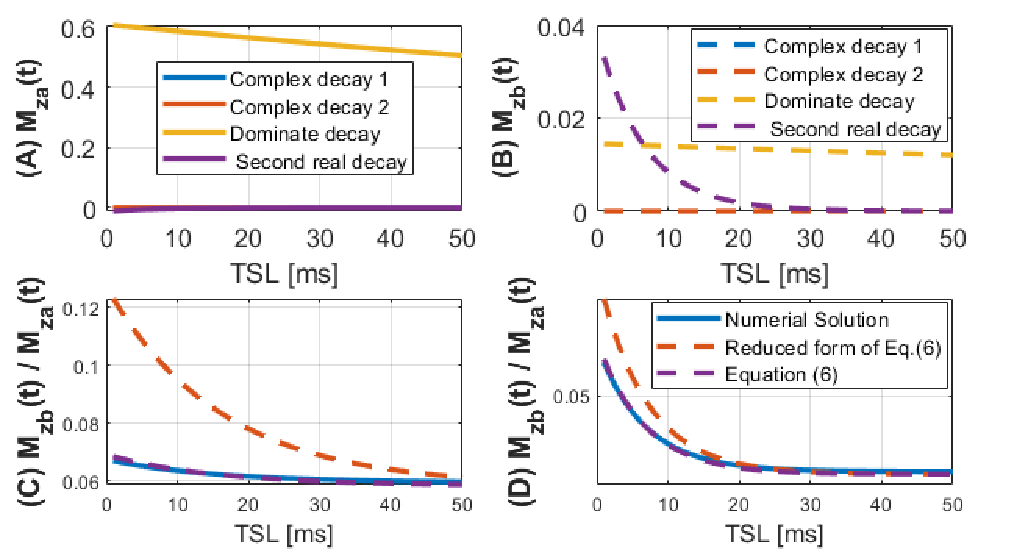}
    \caption{(A) $M_\mathrm{za}(t)$ shows predominantly mono-exponential decay, with a minor secondary component; (B) while $M_\mathrm{zb}(t)$ displays clear bi-exponential behavior. 
    (C,D) Validation of transient state relationship during the spin-lock RF pulse. Comparison of $M_\mathrm{zb}(t)/M_\mathrm{za}(t)$ ratio obtained from three approaches: 1) numerical solution, 2) an reduced form of Eq. \eqref{eqRelation}, see Supporting Information: Derivation S2, and 3) the transient state relationship derived from Eq. \eqref{eqRelation}. (C) correspond to $\omega_\mathrm{1}=2\pi\cdot 350 \text{ rad}$ and $\Delta\omega = 2\pi\cdot3500\text{ rad}$, while (D) corresponds to $\omega_\mathrm{1}=2\pi\cdot 80 \text{ rad}$ and $\Delta\omega=2\pi\cdot800\text{ rad}$. }
    \label{fig:relation}
\end{figure}

\begin{figure}
    \centering
    \includegraphics[width=1\linewidth]{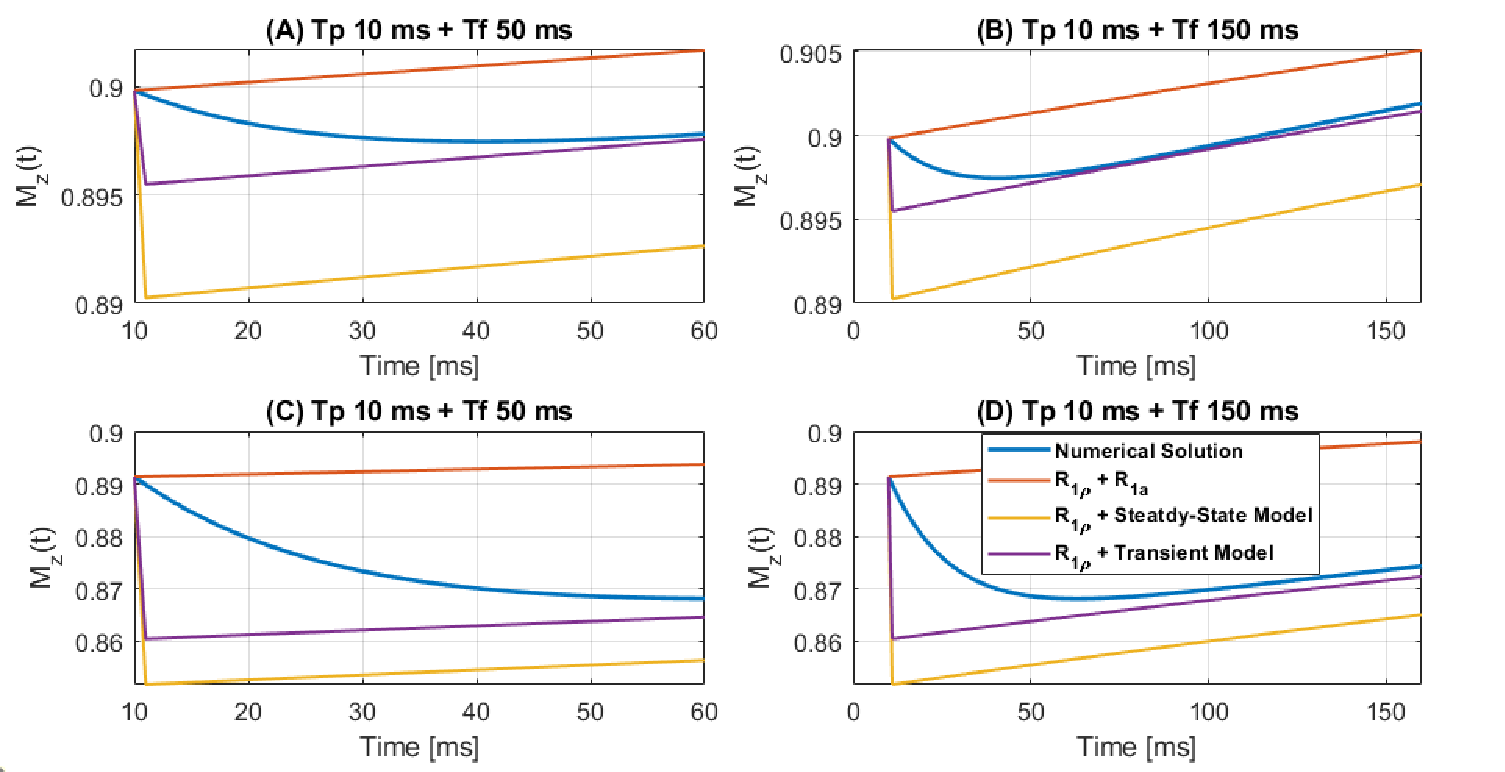}
    \caption{Validation of the analytical solution during free precession periods. Simulations were performed from the end of the last spin-lock period to the end of a free precession period, with varying durations of the free precession stages. (A) and (B) used $\omega_\mathrm{1}=2\pi\cdot 80 \text{ rad}$ and $\Delta\omega=2\pi\cdot 800\text{ rad}$; (C) and (D) used  $\omega_\mathrm{1}=2\pi\cdot 350 \text{ rad}$ and $\Delta\omega=2\pi\cdot 3500\text{ rad}$. In the legend, $R_\mathrm{1\rho}$+$R_\mathrm{1a}$ refers to a model where spin-lock stage is described by mono-exponential decay with rate $R_\mathrm{1\rho}$, and the free precession stage is modeled as $R_\mathrm{1a}$-driven recovery without transferred saturation. The transient state model differs from the steady-state model by assuming that $M_\mathrm{zb}(t)$ and $ M_\mathrm{za}(t)$ do not reach the steady-state at the end of the spin-lock stages. Liver parameters were used in the Bloch-McConnell simulations. }
    \label{fig:pulsed}
\end{figure}

The spin dynamics involving the MT effect during pulsed spin-lock are highly complex, making it challenging to derive a meaningful analytical solution for magnetization evolution. To simplify the analytical solutions, we demonstrate that the transient relationship between $M_\mathrm{za}$ and $M_\mathrm{zb}$ during CW spin-lock can be expressed as follows:
\begin{equation}
   M_\mathrm{zb} = f_{\mathrm{b}}\left(1-\beta\right)M_\mathrm{za},
   \label{eqRelation}
\end{equation}
where
\begin{equation*}
    {\color{black}\beta = \frac{R_{\mathrm{rfb}}}{k_\mathrm{ba}+R_{\mathrm{rfb}}} \times \left( 1-\mathrm{e}^{-\left(R_\mathrm{rfb}+k_\mathrm{ba}\right)T_\mathrm{p}} \right). }
\end{equation*}
{\color{black}The duration of the spin-lock RF pulse, $T_\mathrm{p}$, is around 10 ms in this study}. The detailed derivations of Eq. \eqref{eqRelation} is shown in Supporting Information: Derivation S2.
{\color{black} The validation of the transient relationship of is shown in Fig.\ref{fig:relation}. As depicted, the ratio $M_\mathrm{za}(t)/M_\mathrm{zb}(t)$ derived from the Eq.\eqref{eqRelation} matches well with the numerical solution.}
When $T_\mathrm{p}$ approaches infinity, the steady-state relationship between $M_\mathrm{za}$ and $M_\mathrm{zb}$ can be described with $\beta = {R_{\mathrm{rfb}}}/{\left(k_\mathrm{ba}+R_{\mathrm{rfb}}\right)}$, consistent with the results reported by Roeloffs \emph{et al.} \cite{roeloffs2015ISAR}.
Specifically, the transient relationship described by Eq. \eqref{eqRelation} captures the residual magnetization transfer effects between the two pools as the system evolves from CW spin-lock to free precession. In addition, using the equilibrium magnetization relationship $M_\mathrm{0b} = f_\mathrm{b} M_\mathrm{0a}$, these expressions can be substituted into Eq. \eqref{eqbiexp}, gives: 
\begin{equation}
\begin{split}
M_{\mathrm{za}}(t) &= \left[(1-f_\mathrm{b}\beta)\mathrm{e}^{-R_\mathrm{1a}t} + f_\mathrm{b}\beta\mathrm{e}^{-k_\mathrm{ba}t} \right]M_\mathrm{za}(0)\\ &+ (1-\mathrm{e}^{-R_\mathrm{1a}t})M_\mathrm{0a}.
\label{eqfp1}
\end{split}
\end{equation}
Since $f_\mathrm{b}\ll1$ (with a maximum reported value of $f_\mathrm{b}$ is 19\% in cartilage \cite{zaiss2015combined} and approximately 7\% in the liver) and $\beta<1$, the first decay term, which includes rotational effects, simplifies as $(1-f_\mathrm{b}\beta)\mathrm{e}^{-R_\mathrm{1a}t}\approx \mathrm{e}^{-f_\mathrm{b}\beta-R_\mathrm{1a}t}$. Furthermore, given that $f_\mathrm{b}\beta$ and $\mathrm{e}^{-k_\mathrm{ba}t}$ are negligible under the typical acquisition parameters, i.e., the typical duration of the free precession exceeds tens of milliseconds {\color{black}(set to 50 ms in this study)}, the second exponential decay term is negligible. The resulting equation is expressed as:
\begin{equation}
M_{\mathrm{za}}(t) = M_\mathrm{za}(0)\mathrm{e}^{-f_\mathrm{b}\beta-R_\mathrm{1a}t}+ (1-\mathrm{e}^{-R_\mathrm{1a}t})M_\mathrm{0a}.
\label{eqfp2}
\end{equation}
{\color{black} The validation of Eq.\eqref{eqfp2} is shown in Fig.\ref{fig:pulsed}.}

Free precession is critical in the pulsed spin-lock process, as it allows the system to partially recover from the CW spin-lock. The balance between relaxation and exchange during free precession is critical to the overall efficiency and contrast of the pulsed spin-lock process.
By incorporating the rotational effects, we account for the exchange between \( M_\mathrm{zb} \) and \( M_\mathrm{za} \) during the free precession stage, thereby capturing the magnetization transfer effects specific to this process. 

\subsection{Magnetization Dynamics in Pulsed Spin-Lock Train}
The pulsed spin-lock train is a sequence of repeating individual spin-lock modules consisting of the CW spin-lock and free precession processes. Specifically, in this study, the pulsed spin-lock train comprises $n$ short CW spin-lock RF pulses of duration $T_\mathrm{p}$, uniformly separated by $n-1$ free precession periods, each lasting $T_\mathrm{f}$, as shown in the sequence diagram provided in Supporting Information: Figure S1.

By recursively applying Eq. \eqref{eqcw} and Eq. \eqref{eqfp2}, the magnetization after the entire pulsed spin-lock train can be obtained. To begin, the magnetization at the end of the first spin-lock RF pulse can be expressed as:
\begin{equation}
    M^\mathrm{(1)}=M_{\mathrm{ini}}e^{-R_\mathrm{1\rho}T_\mathrm{p}}+M_\mathrm{ss}(1-e^{-R_\mathrm{1\rho}T_\mathrm{p}}),
\end{equation}
where $M_\mathrm{ini}$ is the initial magnetization before the pulsed sequence. Here $M^\mathrm{(1)}$ serves as the initial condition for the recursive equation.

The recurrence relation relates the magnetization at the end of the $(j-1)$th spin-lock modules $M^\mathrm{(j-1)}$ to the magnetization at the end of the $j$th spin-lock modules $M^\mathrm{(j)}$, can be expressed as:
\begin{equation}
\begin{aligned}
    M^\mathrm{(j)}
    &=\left[M^\mathrm{(j-1)}e^{-R_\mathrm{1a} T_\mathrm{f}-f_\mathrm{b} \beta}+M_\mathrm{0a}(1-e^{-R_\mathrm{1a}T_\mathrm{f}})\right]e^{-R_\mathrm{1\rho}T_\mathrm{p}}\\
    &+M_\mathrm{za}^\mathrm{ss}(1-e^{-R_\mathrm{1\rho}T_\mathrm{p}})\\
\end{aligned} 
\end{equation}

The recursion is subsequently simplified using the formula for a finite geometric series, with the final result shown in Eq.\eqref{eqpulsed1}. 

\begin{figure*}[t!]
\centering
\begin{equation}
\begin{aligned}
    &M^{(n)}=\left(M_{\mathrm{ini}}-M_{\mathrm{ss}}^{(a)}\right) \mathrm{e}^{- \left[(n-1)\cdot \left(R_\mathrm{1a}T_\mathrm{f} +f_\mathrm{b} \beta \right)+n\ \cdot R_\mathrm{1\rho}T_\mathrm{p} \right]}+ M_{\mathrm{ss}}^{(b)},\\
    \textbf{where }&M_{\mathrm{ss}} ^\mathrm{(a)} = \left[ {M_\mathrm{za}^\mathrm{ss} (1-\mathrm{e}^{-R_\mathrm{1\rho}T_\mathrm{p}}) \mathrm{e}^{-\left(R_\mathrm{1a} T_\mathrm{f} +f_\mathrm{b} \beta \right)} + M_\mathrm{0a}(1-\mathrm{e}^{-R_\mathrm{1a}T_\mathrm{f}})}\right]  /\left( {1-e^{-\left(R_\mathrm{1a}T_\mathrm{f} +f_\mathrm{b} \beta +R_\mathrm{1\rho}T_\mathrm{p} \right)}}\right),\\
    &M_{\mathrm{ss}}^\mathrm{(b)} =\left[{M_\mathrm{za}^\mathrm{ss}(1-\mathrm{e}^{-R_\mathrm{1\rho}T_\mathrm{p}}) + M_\mathrm{0a}(1-\mathrm{e}^{-R_\mathrm{1a}T_\mathrm{f}})\mathrm{e}^{-R_\mathrm{1\rho}T_\mathrm{p}}}\right]/\left({1-e^{-\left(R_\mathrm{1a}T_\mathrm{f}+f_\mathrm{b} \beta +R_\mathrm{1\rho}T_\mathrm{p} \right)}}\right).
    \label{eqpulsed1}
    \end{aligned}
\end{equation}
\end{figure*}

We define the inverse duty ratio (IDR) as the ratio of total free precession time to the effective total spin-lock time (\( \mathrm{TSL} = n\times T_\mathrm{p}\)). Mathematically, this is expressed as 
$\mathrm{IDR} = {\left( \left(n-1\right) \times T_\mathrm{f} \right)}/{\left( n \times T_\mathrm{p} \right)}$. Using this definition, the longitudinal relaxation in the rotating frame with pulsed spin-lock, $R_\mathrm{1\rho,pul}$, can be expressed as follows,
\begin{equation}
    R_\mathrm{1\rho,pul}= R_\mathrm{1\rho}+\mathrm{IDR} \cdot R_\mathrm{1a} +\frac{\mathrm{IDR}}{T_\mathrm{f}} \cdot f_\mathrm{b}\beta.
    \label{eqR1rhopul}
\end{equation}
The first term $R_\mathrm{1\rho}$ in Eq. \eqref{eqR1rhopul} is shown in Eq. \eqref{eqR1rho}, which characterizes relaxation during the CW spin-lock. The second term is caused by the $T_\mathrm{1}$ recovery during free precession. The third term characterizes relaxation due to the transferred saturation during the free precession periods.

Equations \eqref{eqpulsed1} and  \eqref{eqR1rhopul} represent a key finding of this work. It shows that, despite the highly complex spin dynamics of magnetization transfer during pulsed spin-lock, the resulting magnetization can be approximated by a simple mono-exponential relaxation model governed by a relaxation rate specific to the pulsed spin-lock, $R_\mathrm{1\rho,pul}$. This will be further validated through simulations and in vivo experiments.

\subsection{MPF-PSL: MPF Mapping Using Pulsed Spin-Lock}
MPF is a key tissue parameter in qMT. A method for MPF mapping using CW spin-lock, termed MPF-SL, was previously introduced for rapid MPF quantification \cite{hou2020mpfsl}. In comparison, MPF mapping based on pulsed spin-lock (MPF-PSL) offers a significantly improved {\color{black}RMP} under typical hardware and SAR constraints. This improvement is crucial for MPF mapping, as the signal specific to macromolecules is inherently small. In MPF-PSL, we measure the difference of $R_\mathrm{1\rho,pul}$ from two acquisitions with the same $\mathrm{IDR}$ and the same duration of free precession $T_\mathrm{f}$:
\begin{equation}
\begin{split}
    &R_\mathrm{mpfsl,pul}= R_\mathrm{1\rho,pul}^\mathrm{(2)}-R_\mathrm{1\rho,pul}^\mathrm{(1)} \\
     &= R_\mathrm{1\rho,pul}(\Delta\omega^\mathrm{(2)},\omega_\mathrm{1}^\mathrm{(2)})
     -R_\mathrm{1\rho,pul}(\Delta\omega^\mathrm{(1)},\omega_\mathrm{1}^\mathrm{(1)})\\
    & = \Delta R_\mathrm{water}+\Delta R_\mathrm{mt,sl} + \frac{\mathrm{IDR}}{T_\mathrm{f}}\Delta R_\mathrm{mt,fp}.
    \label{eqRmpfslpul1}
\end{split}
\end{equation}
Here the superscripts (1) and (2) represent different acquisitions. Note the $R_\mathrm{1a}$-related term in Eq. \eqref{eqR1rhopul} is eliminated in  Eq. \eqref{eqRmpfslpul1}. To further remove the confounding signals from the water pool, similar to the original MPF-SL method \cite{hou2020mpfsl}, we choose the $\Delta\omega^\mathrm{(1)}$, $\omega_\mathrm{1}^\mathrm{(1)}$, $\Delta\omega^\mathrm{(2)}$, $\omega_\mathrm{1}^\mathrm{(2)}$ to satisfy the following condition:
\begin{equation}
    \frac{\Delta\omega^\mathrm{(1)}}{\omega_\mathrm{1} ^\mathrm{(1)}} = \frac{\Delta\omega^\mathrm{(2)}}{\omega_\mathrm{1} ^\mathrm{(2)}}\gg 1.
\label{eqcondition1}
\end{equation}

{\color{black}To optimize the measurement of $R_\mathrm{mpfsl,pul}$, several factors must be carefully considered. Firstly, the frequency offset $\Delta\omega^\mathrm{(1)}$ needs to be sufficiently large to satisfy Eq.~\eqref{eqcondition1} and minimize interference from the water pool. Secondly, $\Delta\omega^\mathrm{(1)}$ should be selected far enough from the chemical exchange pool to avoid contamination from chemical exchange effects, which could otherwise confound the measurement. Third, the choice of $\Delta\omega^\mathrm{(1)}$, $\omega_1^\mathrm{(1)}$, $\Delta\omega^\mathrm{(2)}$, and $\omega_1^\mathrm{(2)}$ directly affects the signal intensity. Since these parameters are required to satisfy Eq.~\ref{eqcondition1}, they can be reduced to three independent variables: $\Delta\omega^{(1)}$, $\omega_1^{(1)}$, and the scaling factor $N$, with the relationships $\Delta\omega^{(2)} = N\Delta\omega^{(1)}$ and $\omega_1^{(2)} = N\omega_1^{(1)}$.
Figure S1(A) in the Supporting Information demonstrates that, given a specific $\Delta\omega^{(1)}$ and $N$, the optimum value of $\omega_1^{(1)}$ can be identified via BM simulation. This approach can be similarly applied to other parameter optimizations.
The values of $\Delta\omega^\mathrm{(1)}$, $\omega_1^\mathrm{(1)}$, $\Delta\omega^\mathrm{(2)}$, and $\omega_1^\mathrm{(2)}$ used in this study are not only distributed very close to the optimal conditions but are also selected within the constraints of hardware limitations.}

{\color{black}Under the conditions defined by Eq. \eqref{eqcondition1}, the relaxation rates $R_\mathrm{1\rho}^{(1)}$ and $R_\mathrm{1\rho}^{(2)}$ are characterized by identical water-pool contributions but exhibit distinct MT-related components. By subtracting $R_\mathrm{1\rho}^{(2)}$ from 
$R_\mathrm{1\rho}^{(1)}$, the water-pool signals are effectively canceled, leaving only the MT-related term $R_{\mathrm{mpfsl,pul}}$. This term, as defined 
by Eq. \eqref{eqRmpfslpul2}, represents a specific component of the MT effect that depends exclusively on MT parameters and is free from contributions arising from the 
water pool.}

To derive Eq. \eqref{eqRmpfslpul2}, we start from the definition of $R_\mathrm{1\rho,pul}$ in Eq. \eqref{eqR1rhopul}. The derivation consists of two components, $R_\mathrm{1\rho}^\mathrm{(2)}-R_\mathrm{1\rho}^\mathrm{(1)}$ and $\mathrm{IDR}\cdot f_\mathrm{b}(\beta^\mathrm{(2)}-\beta^\mathrm{(1)})/T_\mathrm{f}$. The first term follows from the work of Hou \emph{et al.}~\cite{hou2020mpfsl}, while the second term is directly obtained from the expression for $\beta$ in Eq. \eqref{eqRelation}. {\color{black} Comparison of the numerical solution and the value obtained from Eq.\eqref{eqRmpfslpul2} is provided in Fig.\ref{fig:relation2}.}
   
\begin{figure*}[ht]
\centering
\begin{equation}
\begin{split}
    R_{\mathrm{mpfsl,pul}}&= R_\mathrm{1\rho,pul}^\mathrm{(2)}-R_\mathrm{1\rho,pul}^\mathrm{(1)}\\
    &=k_{ba}^2f_b(1+f_b)\left(\frac{1}{(1+f_b)k_{ba}+R_{\mathrm{rfb}}^{(1)}} -\frac{1}{(1+f_b)k_{ba}+R_{\mathrm{rfb}}^{(2)}} \right)\\
    &+\frac{\mathrm{IDR}}{T_\mathrm{f}}\cdot f_{\mathrm{b}}
    \left(\frac{ R_{\mathrm{rfb}}^{(2)}}{k_{ba}+R_{\mathrm{rfb}}^{(2)}}\left(1-\mathrm{e}^{-\left(k_\mathrm{ba}+R_\mathrm{rfb}^{(2)}\right)T_\mathrm{p}}\right)-\frac{ R_{\mathrm{rfb}}^{(1)}}{k_{ba}+R_{\mathrm{rfb}}^{(1)}}\left(1-\mathrm{e}^{-\left(k_\mathrm{ba}+R_\mathrm{rfb}^{(1)}\right)T_\mathrm{p}}\right) \right).
    \label{eqRmpfslpul2}
\end{split}
\end{equation}
\end{figure*}

It is important to notice that Eq. \eqref{eqRmpfslpul2} demonstrates MPF-PSL can be used for qMT imaging without a knowledge of parameters of water pools. This is a significant advantage as the complexity of data acquisition and post-processing increases exponentially with the number of unknown parameters in the model. 

{\color{black} Given that the proposed method is quantitative, MPF maps must be extracted from the measured $R_{\mathrm{mpfsl,pul}}$ data. This is typically accomplished using Eq.~\eqref{eqRmpfslpul2}, which requires knowledge of $B_\mathrm{1}$, $k_\mathrm{ba}$, and $T_\mathrm{2b}$, but not $T_\mathrm{1a}$ or $T_\mathrm{2a}$. Previous studies have shown that $k_\mathrm{ba}$ and $T_\mathrm{2b}$ exhibit minimal variation across subjects\cite{sled2001qMT, underhill2011fast,yarnykh2012fast}, making it reasonable to treat these parameters as constants in MT experiments for specific applications. As a result, published values\cite{stanisz2005parameterSelection} can be adopted without the need for additional measurements. In addition, the $B_1$ map can be rapidly acquired with existing $B_1$ mapping techniques, allowing for accurate calculation of $\omega_\mathrm{1}$ in the presence of significant $B_1$ inhomogeneity.
Alternatively, in more general cases, the MPF can be estimated using a BM equation-based dictionary mapping approach.}

\begin{figure}
    \centering
    \includegraphics[width=1\linewidth]{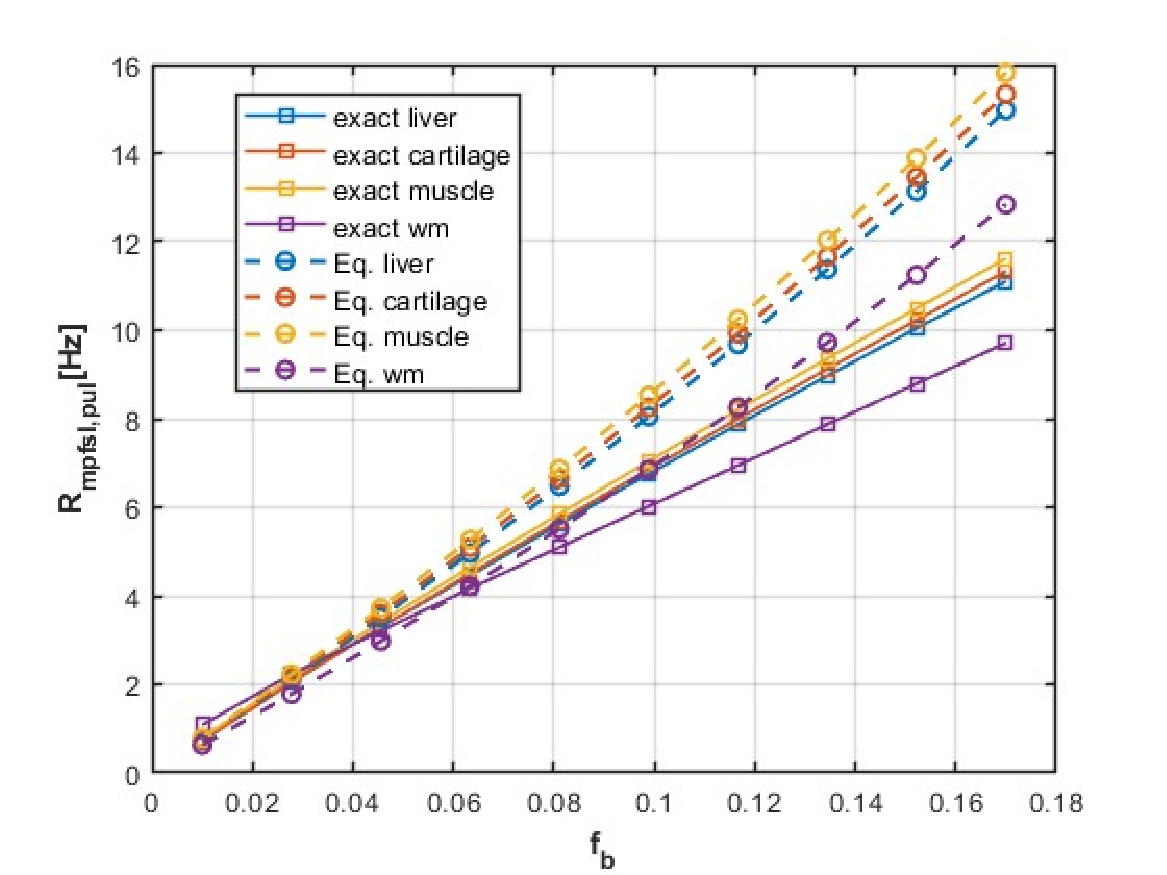}
    \caption{Comparison of the exact $R_\mathrm{mpfsl,pul}$ and the value obtained from Eq. \eqref{eqRmpfslpul2}. The exact $R_\mathrm{mpfsl,pul}$ are calculated by fitting the $R_\mathrm{1\rho}^\mathrm{(1)}$ and $R_\mathrm{1\rho}^\mathrm{(2)}$ separately using Bloch-McConnell simulations. The comparison involves four tissue types: liver, cartilage, muscle, and white matter, based on values reported in previous work\cite{stanisz2005parameterSelection}. The acquisition parameters are set as $\Delta\omega^\mathrm{(1)}=2\pi\cdot 1000 \text{ rad}$, $\Delta\omega^\mathrm{(2)}=2\pi\cdot 4500 \text{ rad}$, $\omega_\mathrm{1}^\mathrm{(1)}=2\pi\cdot 100 \text{ rad}$ and $\omega_\mathrm{1}^\mathrm{(2)}=2\pi\cdot 450 \text{ rad}$. $T_\mathrm{p}$, $T_\mathrm{f}$ and $n$ are 10 ms, 50 ms and 10.}
    \label{fig:relation2}
\end{figure}

\section{Methods} \label{secMethods}
In this section, we first introduce the data acquisition method, followed by the experimental settings for the simulation, phantom, and in vivo studies.

\subsection{The design of data acquisition for MPF-PSL}
\subsubsection{Obtain $R_\mathrm{mpfsl,pul}$ by fitting $R_\mathrm{1\rho,pul}$}
The MT-specific relaxation rate $R_\mathrm{mpfsl,pul}$ can be obtained by subtracting $R_\mathrm{1\rho,pul}^\mathrm{(2)}$ and $R_\mathrm{1\rho,pul}^\mathrm{(1)}$ calculated under the condition specified by Eq. \eqref{eqcondition1}. Note $R_\mathrm{1\rho,pul}$ can be calculated by acquiring multiple pulsed spin-lock weighted images with different numbers of spin-lock modules and fitting the images to the mono-exponential model derived in Eq. \eqref{eqpulsed1}.

\subsubsection{Fast acquisition of $R_\mathrm{mpfsl,pul}$} \label{secAcq}
Calculating $R_\mathrm{mpfsl,pul}$ by fitting $R_\mathrm{1\rho,pul}$ inevitably requires a long scan time. While this is feasible in static tissues such as cartilage and brain, it becomes challenging in dynamic organs like the liver. To avoid the need for individually fitting \( R_\mathrm{1\rho,pul}^\mathrm{(1)} \) and \( R_\mathrm{1\rho,pul}^\mathrm{(2)} \) which results in long scan time, a fast data acquisition process can be employed to directly estimate the MT-specific $R_\mathrm{mpfsl,pul}$ based on the analytical solution of MT during the pulsed spin-lock derived in this study. Specifically, we only collect four images including \(M^\mathrm{(1)}(\Delta\omega^\mathrm{(1)},\omega_\mathrm{1} ^\mathrm{(1)}) \) and \(M^\mathrm{(2)}(\Delta\omega^\mathrm{(1)},\omega_\mathrm{1} ^\mathrm{(1)}) \) at \(\Delta\omega^\mathrm{(1)} \) and \(\omega_\mathrm{1} ^\mathrm{(1)} \) with initial magnetization \( M_\mathrm{ini,1} \) and \( M_\mathrm{ini,2} \), respectively; and \(M^\mathrm{(1)}(\Delta\omega^\mathrm{(2)},\omega_\mathrm{1} ^\mathrm{(2)}) \) and \(M^\mathrm{(2)}(\Delta\omega^\mathrm{(2)},\omega_\mathrm{1} ^\mathrm{(2)}) \) at \(\Delta\omega^\mathrm{(2)} \) and \(\omega_\mathrm{1} ^\mathrm{(2)} \) with initial magnetization \( M_\mathrm{ini,1} \) and \( M_\mathrm{ini,2} \), respectively. Note \( \Delta\omega^\mathrm{(1)}, \omega_\mathrm{1} ^\mathrm{(1)}, \Delta\omega^\mathrm{(2)}, \omega_\mathrm{1} ^\mathrm{(2)} \) satisfy the condition specified in Eq. \eqref{eqcondition1}. Under these conditions, \( R_\mathrm{mpfsl,pul} \) can be calculated from these four images using Eq. \eqref{eqacq}. The fast acquisition approach enables MPF-PSL imaging of moving organs using breath-hold. In the following, $R_\mathrm{mpfsl,pul}$ will be obtained using this method unless noted.

\begin{figure*}[ht]
\centering
\begin{equation}
\begin{split}
    R_\mathrm{mpfsl,pul} & = R_\mathrm{1\rho,pul}^\mathrm{(2)} -  R_\mathrm{1\rho,pul}^\mathrm{(1)}\\ & \approx -\mathrm{log}\left(\frac{ M^\mathrm{(1)}(\Delta\omega^\mathrm{(1)},\omega_\mathrm{1} ^\mathrm{(1)}) -M^\mathrm{(2)}(\Delta\omega^\mathrm{(1)},\omega_\mathrm{1} ^\mathrm{(1)}) } {M^\mathrm{(1)}(\Delta\omega^\mathrm{(2)},\omega_\mathrm{1} ^\mathrm{(2)}) - M^\mathrm{(2)}(\Delta\omega^\mathrm{(2)},\omega_\mathrm{1} ^\mathrm{(2)}) } \right)/\mathrm{TSL},
    \label{eqacq}
\end{split}
\end{equation}
\end{figure*}

Note the change of initial magnetization can be realized in multiple ways. One common technique involves using one or more RF pulses prior to the spin-lock pulses. These RF pulses offer flexibility in terms of flip angle, phase, frequency modulation, duration, amplitude, etc. For example, we use an adiabatic 180$^{\circ}$ inversion toggling RF pulse\cite{jin2012toggling, jiang2019toggling, mangia2011toggling} to create a different initial magnetization. 

\subsection{Simulation studies}

\subsubsection{Simulation study 1: Full-equation Bloch-McConnell simulation of $R_\mathrm{mpfsl,pul}$}
A comprehensive two-pool BM simulation was performed to evaluate the sensitivity of $R_\mathrm{mpfsl,pul}$ to the parameters in the MT model. The parameters of the liver tissue were used in this study, including: $T_\mathrm{1a}=812 \text{ ms}$, $T_\mathrm{2a}=42 \text{ ms}$, $T_\mathrm{2b}=7.7 \mu\text{s}$, $f_\mathrm{b}=6.9\% $ and $k_\mathrm{ba} = 51 \text{s}^\mathrm{-1}$, based on the values reported in a prior study \cite{stanisz2005parameterSelection}. During simulation, only one of these parameters varied within a typical range while all the other parameters were kept unchanged by using the aforementioned values. The simulation assumes that $R_\mathrm{1a}=R_\mathrm{1b}$, as discussed in previous studies\cite{yarnykh2002pulsed, yarnykh2012fast}. Using the acquisition method detailed in Section \ref{secAcq}, $R_\mathrm{mpfsl,pul}$ was calculated based on the resulting longitudinal magnetization. 
The sequence parameters include $\Delta\omega^\mathrm{(1)}=2\pi\cdot 800\text{ rad}$, $\Delta\omega^\mathrm{(2)}=2\pi\cdot 3500\text{ rad}$, $\omega_\mathrm{1}^\mathrm{(1)}=2\pi\cdot 80\text{ rad}$ and $\omega_\mathrm{1}^\mathrm{(2)}=2\pi\cdot 350\text{ rad}$.
The performance of MPF-PSL was tested using various combinations of acquisition parameters of $[T_\mathrm{p}, T_\mathrm{f}, n]$ including: [5\text{ ms}, 50\text{ ms}, 20], [10\text{ ms}, 50\text{ ms}, 10] and [20\text{ ms}, 50\text{ ms}, 5]. We also performed simulation of the original MPF-SL \cite{hou2020mpfsl} using the same parameters. The same TSL of 100 \text{ms} were used in all simulations. Note that such a long TSL for the original MPF-SL is generally challenging, particularly when using body RF transmit, due to hardware limitations. All the simulations, along with the reconstruction of parameter maps and image analysis, were implemented in MATLAB R2024a (MathWorks, USA). The code supporting the findings of this study will be made publicly available upon the acceptance of this manuscript for publication.

\begin{figure}
    \centering
    \includegraphics[width=1\linewidth]{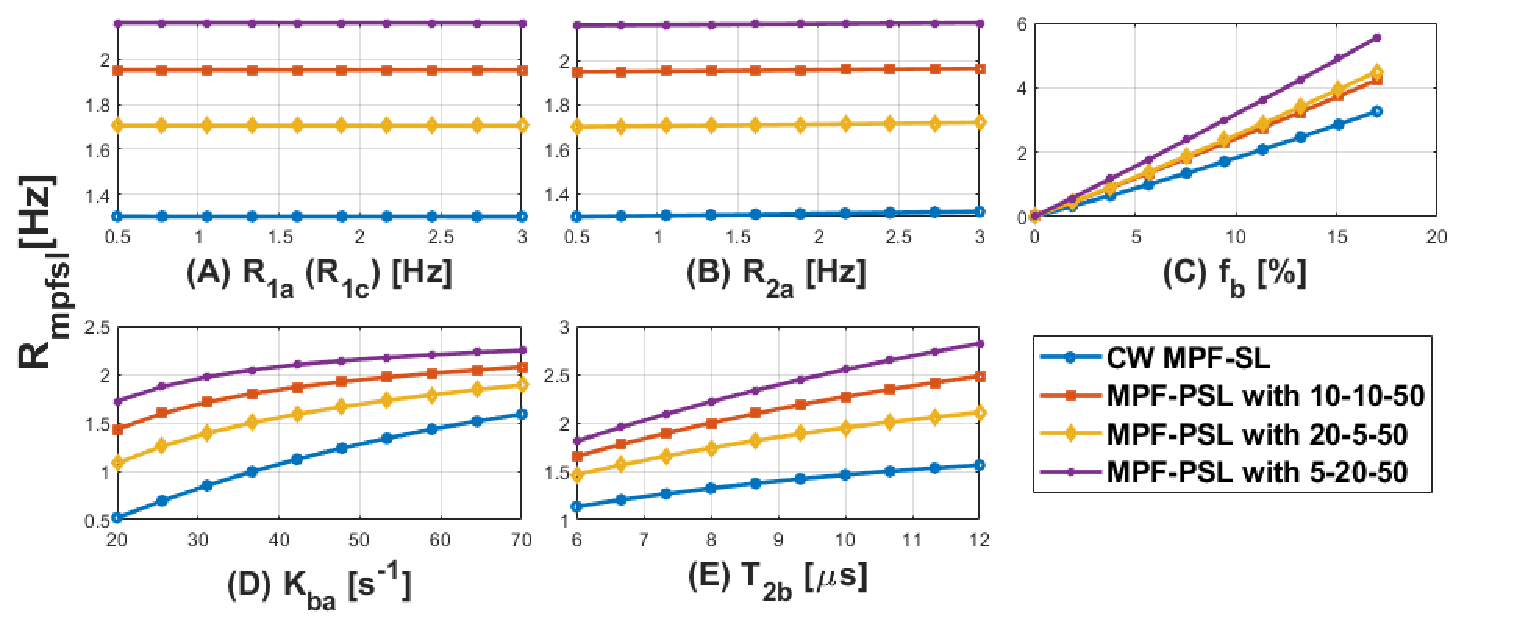}
    \caption{$R_\mathrm{mpfsl,pul}$ and $R_\mathrm{mpfsl}$ sensitivity to tissue parameters. Liver parameters were used in this full BM simulation study. The variations of parameters fall within the typical ranges. The MPF-PSL methods utilize train pulses with different parameter settings ($T_\mathrm{p}-n-T_\mathrm{f}$).}
    \label{fig:sim1}
\end{figure}

\begin{figure}
    \centering
    \includegraphics[width=1\linewidth]{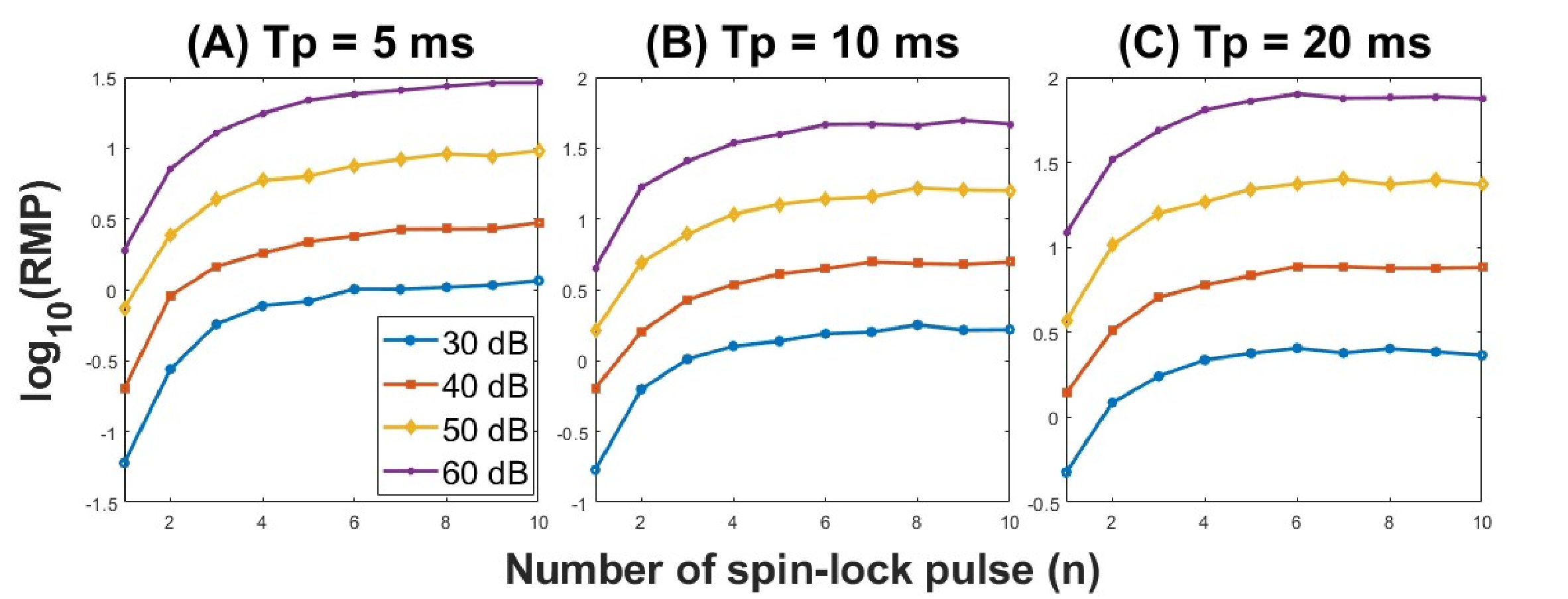}
    \caption{Improved {\color{black}RMP} using spin-lock train sequence in $R_\mathrm{mpfsl,pul}$. The noise is introduced into the $M_\mathrm{z}$ signal according to varied noise levels indicated in the image legend. The simulation was based on liver parameters. Across figures (A), (B), and (C), a constant $T_\mathrm{f}$ of 50ms is maintained, with $T_\mathrm{p}$ differing as 5ms, 10ms, and 20ms, respectively. The results show that an increase in the number of spin-lock units leads to an improvement in the {\color{black}RMP} of $R_\mathrm{mpfsl,pul}$.}
    \label{fig:sim2}
\end{figure}

\subsubsection{Simulation study 2: The {\color{black} RMP} gained using MPF-PSL}
This simulation aimed to demonstrate that the {\color{black} precision and reliability of the measurement} improves with an increasing number of spin-lock modules in MPF-PSL, corresponding to a longer total spin-lock duration. {\color{black}To evaluate the relative consistency of the measured values, we used the Relative Measurement Precision (RMP), which was calculated using the following equation:
\begin{equation}
    {\color{black}\mathrm{RMP}}=\frac{\mu_{R_\mathrm{mpfsl,pul}}}{\sigma_{R_\mathrm{mpfsl,pul}}},
    \label{eqsnr}
\end{equation}
where the $\mu_{R_\mathrm{mpfsl,pul}}$ represents the mean of $R_\mathrm{mpfsl,pul}$ and $\sigma_{R_\mathrm{mpfsl,pul}}$ denotes the standard deviation of $R_\mathrm{mpfsl,pul}$.}
 
Here, we employed varying $T_\mathrm{p}$ with 5 ms, 10 ms and 20 ms, while maintaining a fixed $T_\mathrm{f}$ of 50 ms. Gaussian noise of zero mean \cite{chen2015errors} at levels of 30 dB, 40 dB, 50 dB, and 60 dB was introduced to the resulting longitudinal magnetization signals. We then calculated the $R_\mathrm{mpfsl}$ {\color{black}and the corresponding RMP} as the number of spin-lock modules ($n$) increased.

\subsection{Phantom and in vivo studies}
All data were acquired using a Philips Elition 3T MRI scanner (Philips Healthcare, the Netherlands). The in vivo study was conducted with the approval of the institutional review board. An 8-channel head coil, a 16-channel T/R knee coil, and a 32-channel cardiac coil (Invivo Corp, Gainesville, FL, USA) were used for the phantom, knee and liver study, respectively. Body RF transmit was used for phantom and liver scan. 
Pulsed spin-lock preparation was performed using the aforementioned fast acquisition approach. Following the pulsed spin-lock preparation, imaging data were acquired using a 2D fast/turbo spin echo (FSE/TSE) pulse sequence. The acquisition parameters $T_\mathrm{p}=10\text{ ms}$, $T_\mathrm{f}=50\text{ ms}$, $n=10$ were used throughout all phantom and in vivo studies.

\subsubsection{Phantom studies 1: The relations between $R_\mathrm{mpfsl,pul}$ and MPF}
In this study, we investigated the relationship between the measured $R_\mathrm{mpfsl,pul}$ and MPF. Phantoms with agarose concentrations 1\% to 5\% were prepared. 
The imaging parameters were as follows: field of view (FOV) 250 mm $\times$ 250 mm, resolution 2 mm $\times$ 2 mm, {\color{black} single slice,} slice thickness 7 mm, repetition time/echo time (TR/TE) 3000/17 ms. The sequence parameters were set as $\Delta\omega^\mathrm{(1)}=2\pi\cdot 1000\text{ rad}$, $\Delta\omega^\mathrm{(2)}=2\pi\cdot 4500\text{ rad}$, $\omega_\mathrm{1}^\mathrm{(1)}=2\pi\cdot 100\text{ rad}$ and $\omega_\mathrm{1}^\mathrm{(2)}=2\pi\cdot 450\text{ rad}$.

\begin{figure}
    \centering
    \includegraphics[width=1\linewidth]{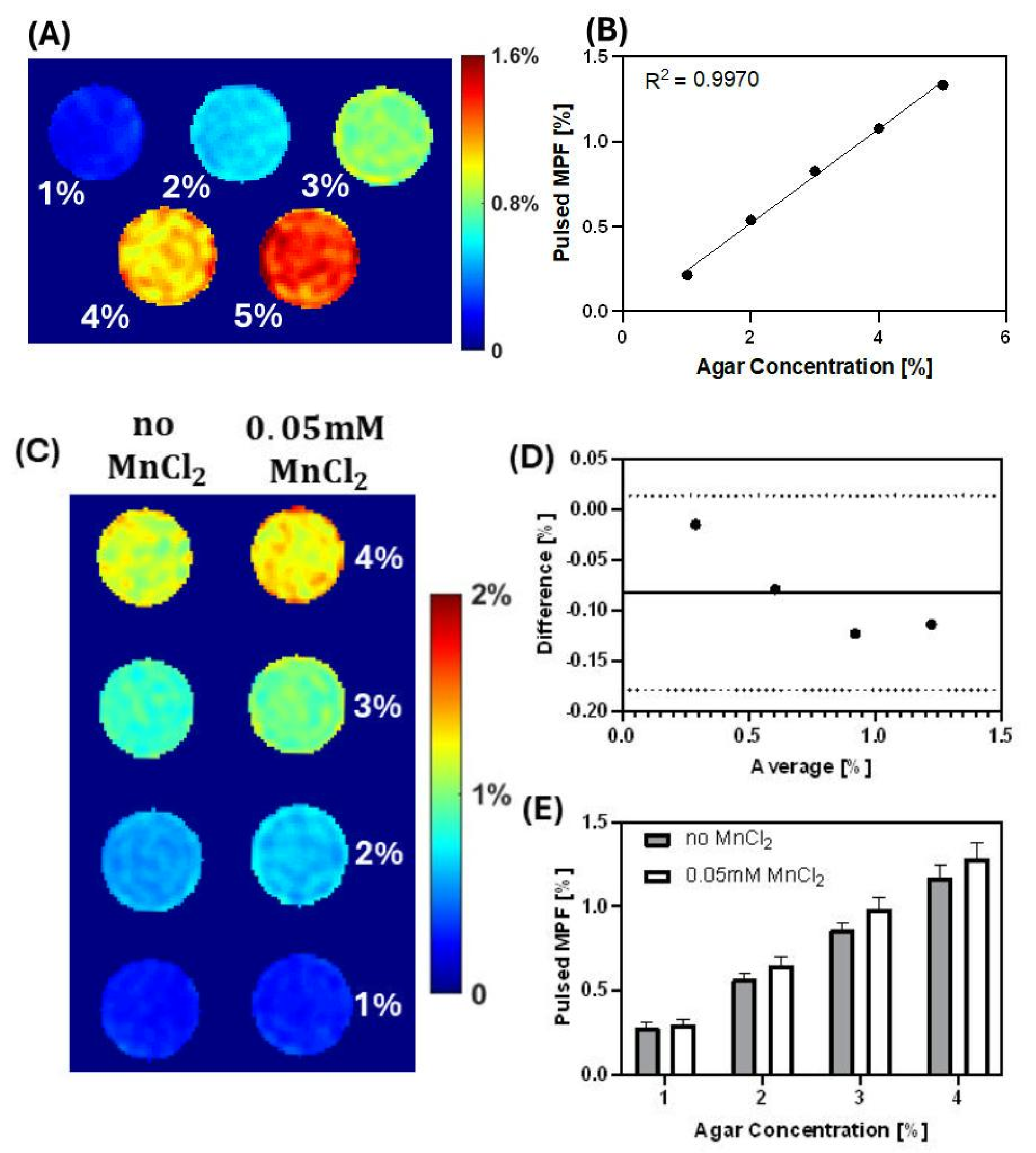}
    \caption{{\color{black}Phantom Study 1 and 2. (A) The left is the map of 1\% to 5\% agarose phantoms with MPF-PSL. (B) On the right, the scatter plots represent the linear regression between agarose concentration and the corresponding measured MPF values.(C) MPF map of 1\% to 4\% agarose phantms obtained using MPF-PSL methods. (D) Bland-Altman plot comparing MPF results with and without $\text{MnCl}_2$. The solid line represents the average difference, which is -0.1\%, while the dotted lines denote the 95\% limits of agreement, spnning from -0.18\% to 0.01\%. (E) Comparison of pulsed MPF measurements for phantoms with and without 0.05 mM $\text{MnCl}_2$}}
    \label{fig:phan1}
\end{figure}

\subsubsection{Phantom studies 2: Insensitivity of MPF-PSL to free water pool}
To validate the insensitivity of MPF-PSL to the free water pool, we prepared two groups of agarose phantoms, each containing agarose at concentrations of 1\% to 4\%. In one group, 0.05 mM $\text{MnCl}_2$ was added, while the other group contained no $\text{MnCl}_2$. The addition of $\text{MnCl}_2$ served to modulate $R_\mathrm{1a}$ and $R_\mathrm{2a}$, the relaxation rates of the free water pool. The key hypothesis was that $R_\mathrm{mpfsl,pul}$ would remain consistent between the two groups for the same agarose concentration, thereby confirming the method's insensitivity to variations in free water pool parameters. Imaging parameters and sequence parameters for this experiment were identical to those used in phantom study 1.

\subsubsection{Phantom studies 3: Improved {\color{black}RMP} with increasing number of spin-lock modules}
A phantom with a 2\% agarose concentration and a diameter of 10 cm was prepared to validate the relationship between the number of spin-lock modules ($n$) and the {\color{black}RMP} of $R_\mathrm{mpfsl,pul}$. In this study, $T_\mathrm{p}$ and $T_\mathrm{f}$ were fixed at 10 ms and 50 ms, respectively, while $n$ was increased from 1 to 10. The imaging parameters for this experiment were the same as those used in phantom study 1. The sequence parameters were set as $\Delta\omega^\mathrm{(1)}=2\pi\cdot 800\text{ rad}$, $\Delta\omega^\mathrm{(2)}=2\pi\cdot 3500\text{ rad}$, $\omega_\mathrm{1}^\mathrm{(1)}=2\pi\cdot 80\text{ rad}$ and $\omega_\mathrm{1}^\mathrm{(2)}=2\pi\cdot 350\text{ rad}$.
To evaluate the {\color{black}RMP}, a region of interest (ROI) was manually drawn to cover at least 80\% of the phantom. 

\begin{figure*}
    \centering
    \includegraphics[width=1\linewidth]{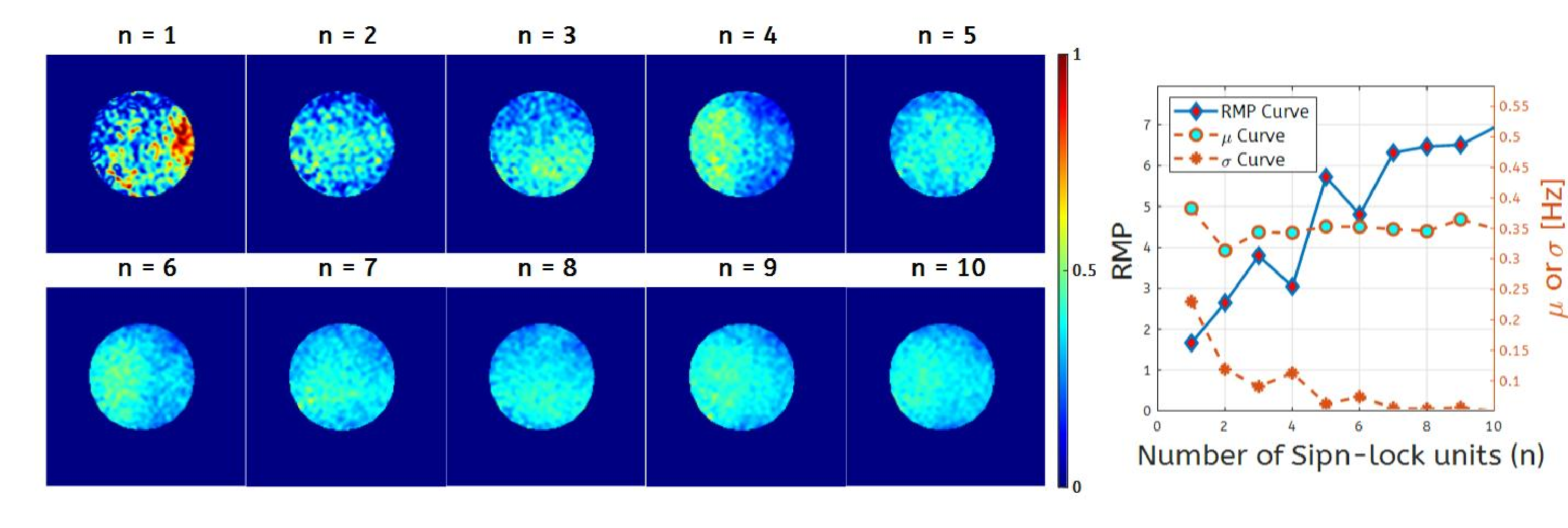}
    \caption{Left: $R_\mathrm{mpfsl,pul}$ images of phantoms with different n. Right: Corresponding {\color{black}RMP}, $\mu$, and $\sigma$ curves, showing {\color{black}RMP} increases with spin-lock units.}
    \label{fig:phan3}
\end{figure*}

\subsubsection{In vivo study 1: Validation of relaxation model of MPF-PSL}
One main finding of this work is that magnetization prepared using pulsed spin-lock can be characterized by a mono-exponential relaxation model as shown in \eqref{eqpulsed1}. In this in vivo study, we validate this in knee cartilage scan. We collected 12 pulsed spin-lock prepared images with varying numbers of spin-lock modules (n=1, 2, 4, 5, 7, 9, 12, 15, 25, 29, 30, and 40) and investigated whether their signal decay follows mono-exponential model. Two sets of images were acquired: one with lower RF ($\omega_\mathrm{1}=80\cdot2\pi \text{ rad}$ and $\Delta\omega=800\cdot 2 \pi \text{ rad}$) and the other with higher ($\omega_\mathrm{1}=350\cdot2\pi \text{ rad}$ and $\Delta\omega=3500\cdot 2 \pi \text{ rad}$) RF. Additionally, $ T_\mathrm{p}$ was set to 10 ms and $T_\mathrm{f}$ was set to 50 ms. The spectral pre-saturation with inversion recovery (SPIR) was applied to suppress fat signals. The imaging parameters include: FOV 160 mm $\times$ 160 mm, TR/TE 15000/11 ms, resolution 1.5 mm $\times$ 1.5 mm{\color{black}, single slice}.

\subsubsection{In vivo studies 2: MPF mapping of liver fibrosis using MPF-PSL}
One promising application of MPF mapping is assessment of liver fibrosis. Note the progression of fibrosis is characterized by increasing deposition of extracellular matrix components, particularly collagen, in fibrotic tissues, which is expected to reflect in MPF mapping. In this study, we recruited 6 patients with metabolic dysfunction-associated steatotic liver disease (MASLD) with different stages of fibrosis. The fibrosis stage was confirmed by liver biopsy. This study included two patients with early-stage liver fibrosis (F1/2), two patients with advanced fibrosis (F3), and two patients with cirrhosis (F4). The imaging parameters include: FOV 380 mm $\times$ 380 mm, TR/TE 3000/13 ms, resolution 2 mm $\times$ 2 mm, slice thickness 7 mm, number of slices 6. Double inversion recovery (DIR) was used to suppress blood signals, and SPIR was applied to suppress fat signals. The other parameters are the same as those in simulation study 1. 

\begin{figure}
    \centering
    \includegraphics[width=1\linewidth]{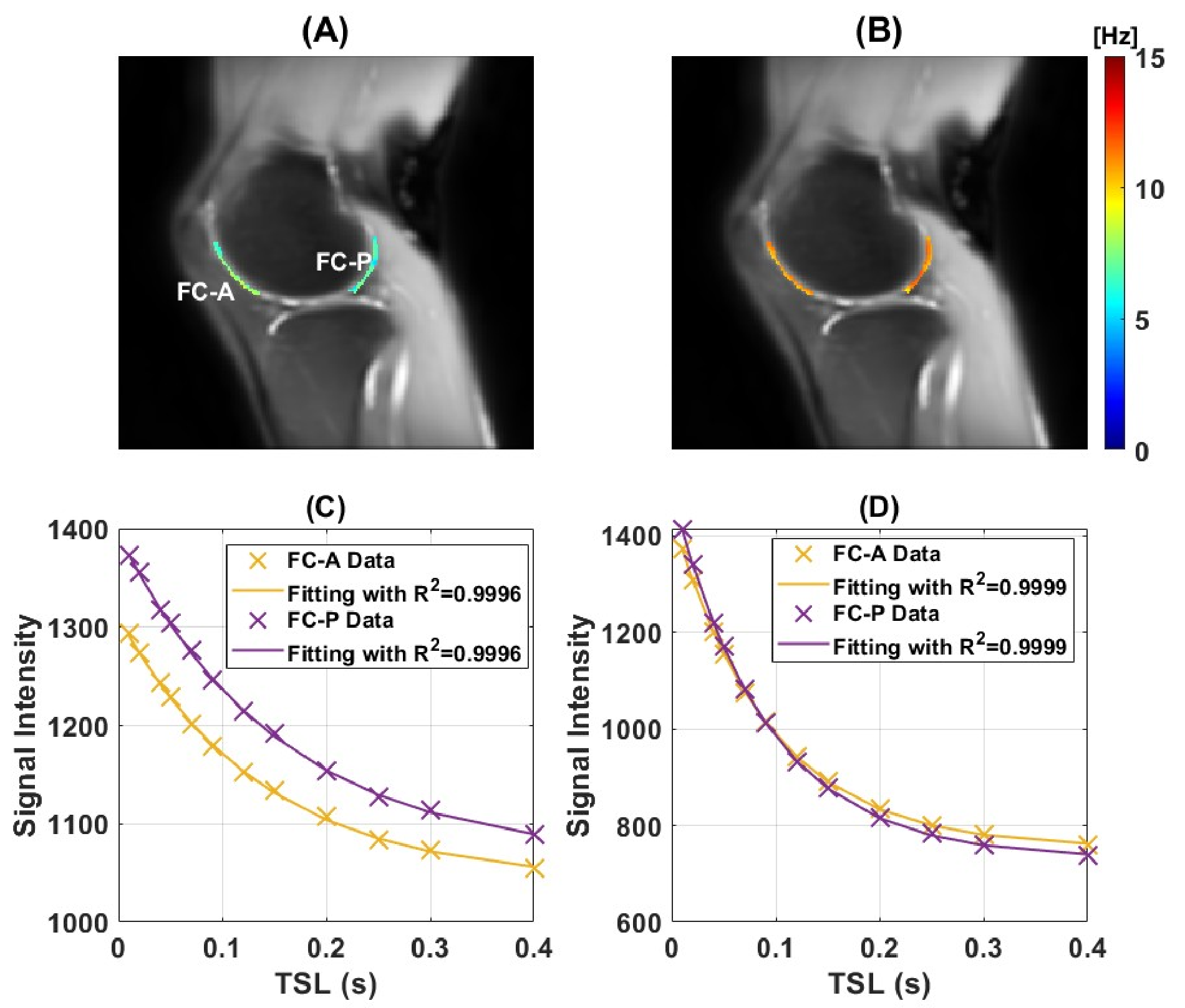}
    \caption{In vivo knee study of a healthy volunteer for off-resonance pulsed $R_\mathrm{1\rho}$ fitting. ROIs FC-A, FC-P correspond to the anterior, and posterior regions of the femoral cartilage respectively. (A) Low RF pulse setting: $\omega_\mathrm{1} = 80 \cdot 2\pi \text{ rad}$ and $\Delta\omega = 800 \cdot 2\pi \text{ rad}$; (B) High RF pulse setting: $\omega_\mathrm{1} = 350 \cdot 2\pi \text{ rad}$ and $\Delta\omega = 3500 \cdot 2\pi \text{ rad}$. (C,D) Low- and high-RF cartilage signals for the 2 ROIs were individually fitted to a mono-exponential model, {\color{black}with the goodness-of-fit metric ($R^2$) embedded in the legend for quantitative assessment.} The fitting results for other ROIs are provided in Supporting Information: Figure S{\color{black}2}.}
    \label{fig:knee}
\end{figure}

\begin{figure}
    \centering
    \includegraphics[width=1\linewidth]{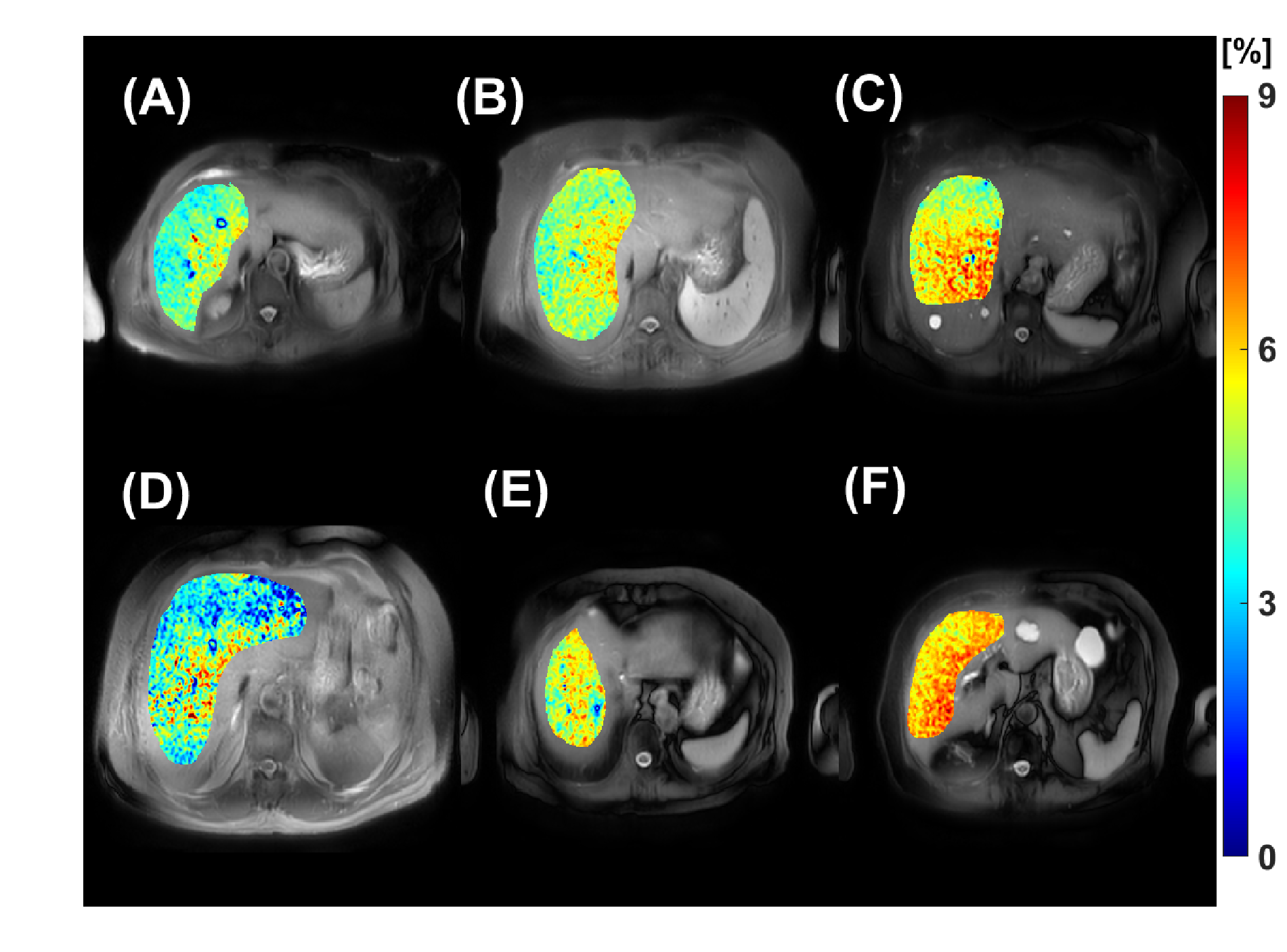}
    \caption{In vivo study of six patients with MASLD, with different stages of fibrosis confirmed by liver biopsy. (A) and (D): two patients with early-stage liver fibrosis (F1/2); (B) and (E): advanced fibrosis (F3); (C) and (F): cirrhosis (F4). Note the increase in MPF in the fibrotic liver, reflecting the deposition of collagen in fibrotic tissues.}
    \label{fig:invivo}
\end{figure}

\section{Results}\label{secResults}

Figure \ref{fig:sim1} presents the results from simulation study 1. The proposed MPF-PSL method demonstrates high sensitivity to MPF while exhibiting negligible sensitivity to water pool parameters such as $R_\mathrm{1a}$ and $R_\mathrm{2a}$. {\color{black} 
It should also be noted that MPF-PSL is somewhat sensitive to $k_\mathrm{ba}$ and $T_\mathrm{2b}$. However, as discussed in Section 2.5, both parameters exhibit limited variability in the certain application \cite{sled2001qMT, underhill2011fast, yarnykh2012fast}, allowing the use of established literature values in practice.}

Figure \ref{fig:sim2} illustrates the results of simulation study 2. The {\color{black}RMP} of $R_\mathrm{mpfsl,pul}$ is shown to increase with total spin-lock time across various noise levels and $T_\mathrm{p}$ values. This highlights the advantage of the proposed MPF-PSL approach. Unlike the original MPF-SL method, which is limited by hardware constraints and unable to sustain prolonged CW spin-lock pulses, the MPF-PSL approach allows for extended spin-lock durations. This capability results in higher {\color{black}RMP, reflecting the improved reliability, consistency, and accuracy of the measurements, which are critical for in vivo imaging.}

Figure \ref{fig:phan1}{\color{black}(A) and (B) show} the results from phantom study 1. {\color{black} A strong linear relationship ($R^2=0.9970$) is observed between the measured MPF and agarose concentration. This demonstrates the effectiveness of the proposed MPF-PSL method in detecting changes in MPF.}

Figure \ref{fig:phan1}{\color{black} (C)-(E) present} the results from phantom study 2. The MPF-PSL method achieves consistent quantitative measurements independent of $\text{MnCl}_2$ concentration, provided the agarose concentration remains constant.  
{\color{black} Since the $\text{MnCl}_2$ significantly alters $R_\mathrm{1a}$ and $R_\mathrm{2a}$ of the phantom, as demonstrated in Supporting Information: Tables S1}, these results further confirm the insensitivity of MPF-PSL to variations in water pool parameters.

Figure \ref{fig:phan3} depicts the results from phantom study 3, demonstrating a significant {\color{black}RMP} gain in $R_\mathrm{mpfsl,pul}$ with an increasing number of spin-lock modules. The $R_\mathrm{mpfsl,pul}$ images indicate that as the number of spin-lock modules increases, image uniformity improves. The $\mu$ curve on the right shows the measured $R_\mathrm{mpfsl,pul}$ values, revealing that with fewer spin-lock modules, the measured values are artificially elevated and less reliable due to high noise introducing non-zero biases to the true $R_\mathrm{mpfsl}$ values. {\color{black} The $\sigma$ curve decreases as the number of spin-lock units increases, reflecting reduced data fluctuations and improved consistency in the measurement results.}

Figure \ref{fig:knee} shows the results from the knee scan. {\color{black}In our analysis, all $R^2$ values exceeded 0.99, indicating that the signal from cartilage is well-fitted by a mono-exponential model, validating that the magnetization under the proposed MPF-PSL acquisition can be adequately modeled by a mono-exponential function with acceptable error. Additionally, as shown in the figure, the fitting curves align closely with the data points, further supporting this conclusion.}

Finally, Fig.\ref{fig:invivo} presents results from patients with liver fibrosis. The proposed MPF-PSL method successfully detects elevated MPF levels in fibrotic liver tissue, likely due to the deposition of collagen in liver tissues during fibrosis.

\section{Discussion} \label{secDiscussion}

\begin{figure*}[ht]
\centering
\begin{equation}
\begin{split}
    &R_{\mathrm{mpfsl,pul}}= k_{ba}^2f_b(1+f_b)\left(\frac{1}{(1+f_b)k_{ba}+R_{\mathrm{rfb}}^{(1)}+{\color{blue}R_\mathrm{1b}}} -\frac{1}{(1+f_b)k_{ba}+R_{\mathrm{rfb}}^{(2)}+{\color{blue}R_\mathrm{1b}} } \right)\\
    &+\frac{\mathrm{IDR}}{T_\mathrm{f}}\cdot f_{\mathrm{b}}
    \left(\frac{ R_{\mathrm{rfb}}^{(2)}}{k_{ba}+R_{\mathrm{rfb}}^{(2)}+{\color{blue}R_\mathrm{1b}}}\left(1-\mathrm{e}^{-\left(k_\mathrm{ba}+R_\mathrm{rfb}^{(2)}+{\color{blue}R_\mathrm{1b}}\right)T_\mathrm{p}}\right)-\frac{ R_{\mathrm{rfb}}^{(1)}}{k_{ba}+R_{\mathrm{rfb}}^{(1)}+{\color{blue}R_\mathrm{1b}}}\left(1-\mathrm{e}^{-\left(k_\mathrm{ba}+R_\mathrm{rfb}^{(1)}+{\color{blue}R_\mathrm{1b}}\right)T_\mathrm{p}}\right) \right).
    \label{eqt1b}
\end{split}
\end{equation}
\end{figure*}

Imaging of macromolecule levels holds substantial clinical value but necessitates both reliable quantification and efficient scan protocols to enable routine application. Traditional qMT models incorporate parameters from both the water and MT pools, while the spin-lock-based approaches, referred to as qMT-SL methods, simplifies quantification by focusing exclusively on MT parameters and minimizing the direct water saturation effects. However, given the low concentration of macromolecules in tissues, enhancing signal strength in qMT-SL methods is critical to achieving reliable quantification.

Using relatively long total spin-lock durations is advantageous for qMT-SL for several reasons. 
First, qMT-SL relies on relaxometry analysis, where accurate parameter estimation requires sufficient magnetization decay during the spin-lock process. If the spin-lock duration is too short, the decay is minimal, resulting in unreliable parameter estimation. By extending the spin-lock duration, the robustness of relaxation parameter estimation can be significantly enhanced, as demonstrated in previous studies \cite{huang2022uncertainty}. Second, longer spin-lock durations inherently improve signal fidelity by mitigating modeling errors in our mono-exponential approximation of CW spin-lock magnetization dynamics. This simplified model assumes that the magnetization dynamics follow a mono-exponential behavior with a constant scaling factor across different acquisition parameters. However, further mathematical analysis (provided in Supporting Information: Derivation S3) reveals that this approximation neglects secondary terms in the exact solution. Consequently, \(  R_{\mathrm{mpfsl,pul}}\) increases with longer total spin-lock durations. Notably, as the spin-lock duration is extended, the errors introduced by these assumption violations diminish, and \(  R_{\mathrm{mpfsl,pul}}\) converges to a constant value. Thus, extended spin-lock durations enhance robustness and improve signal levels in qMT-SL.

There is ongoing debate regarding the values of $R_\mathrm{1a}$ and $R_\mathrm{1b}$. Most studies assume that both rates are around 1 Hz, often fixing $R_\mathrm{1b}=1$ Hz or setting $R_\mathrm{1b}=R_\mathrm{1a}$ \cite{yarnykh2002pulsed, dortch2011assum}, as we did in simulation study 1. However, recent studies have reported substantially different longitudinal relaxation times of the free and semi-solid spin pools, such as $T_\mathrm{1a} \approx 2\ \text{s}$ and $T_\mathrm{1b} \approx 0.3\ \text{s}$ in white matter at 3T\cite{asslander2024unconstrained, van2016wt, helms2009wt, manning2021wt, samsonov2021wt}.
In this case, the assumption $R_\mathrm{1a}=R_\mathrm{1b}$ is no longer valid.
This can be addressed by assuming $R_\mathrm{1a} \ll R_\mathrm{b}$ and including a water pool-independent parameter, $R_\mathrm{1b}$, in the original framework, as shown in Eq. \eqref{eqt1b}.

Quantitative MT imaging has been extensively studied for myelin imaging and has recently gained attention for assessing liver fibrosis \cite{yarnykh2015liver, hou2023detecting, wilczynski2023MEX}. Liver fibrosis, characterized by collagen deposition in the extracellular matrix, involves macromolecules such as collagen that exhibit a MT effect. In this study, we demonstrated that the proposed MPF-PSL approach holds potential for differentiating various stages of liver fibrosis. On the same MRI system, the original MPF-SL method was unable to achieve sufficient {\color{black}RMP} in liver scans due to strict RF hardware constraints. In contrast, the proposed MPF-PSL achieved substantially higher {\color{black}RMP}, highlighting its advantages in liver imaging.

Liver imaging poses unique challenges due to the complex physiological conditions of the liver, including fat deposition, blood flow, iron deposition, and respiratory motion. Additionally, significant $B_\mathrm{1}$ RF and $B_\mathrm{0}$ field inhomogeneity further complicate imaging. Since detecting subtle signals from collagen in the liver is the primary goal, it is essential to not only enhance the signal from the MT pool but also suppress confounding signals caused by physiological factors and system imperfections. The qMT-SL approach may address these challenges effectively, as it can be combined with blood suppression \cite{chen2016breath} and fat suppression \cite{hou2023detecting} modules, enabling data acquisition within a brief breath-hold. Furthermore, qMT-SL has been shown to be insensitive to iron deposition and robustness against $B_\mathrm{1}$ RF and $B_\mathrm{0}$ field inhomogeneity \cite{hou2020mpfsl,hou2023detecting}, making it promising for liver fibrosis imaging.

This work has some limitations. 
First, we introduced additional pulse sequence parameters $T_\mathrm{p}$, $T_\mathrm{f}$ and $n$. In this study, the pulse duration $T_\mathrm{p}$ and the gap between pulses $T_\mathrm{f}$ were fixed. However, these parameters do not necessarily need to remain constant during scanning. Further work is needed to optimize these pulse sequence parameters to achieve the best performance. 
Second, as in previous studies, the proposed MPF-PSL method assumes constant exchange and relaxation rates for the MT pool. While this assumption may be valid in the liver, variations in these parameters across other tissues could introduce bias, making the resulting maps more appropriately described as MPF-weighted. Since obtaining ground truth MPF values in vivo is challenging, the focus shifts toward assessing the parameter's diagnostic utility in clinical applications. Notably, the relaxation rate $R_\mathrm{mpfsl,pul}$ shows strong clinical potential in our studies.
{\color{black} Third, despite preliminary clinical studies conducted on the conventional MPF-SL technology for detecting early-stage liver fibrosis\cite{hou2023detecting}, comprehensive validation of the proposed MPF-PSL technology requires a larger-scale patient study and rigorous statistical analysis of MPF differences across groups.}
Finally, in this work, we demonstrated the proposed method by MPF mapping, as MPF is the parameter of primary interest in most applications. While our theoretical framework indicates that the proposed method has potential for quantifying other MT parameters, further studies are needed to extend this approach for comprehensive qMT imaging.

\section{Conclusion} \label{secConclusion}
We proposed a pulsed spin-lock approach to address hardware limitations of clinical MRI systems for quantitative MT imaging based on spin-lock MRI. An analytical model of the spin dynamics underlying the proposed approach was developed by segmenting the sequence into spin-lock and free precession stages, connected through a transient state relationship. The theoretical framework demonstrates that the proposed approach minimizes confounding effects from the water pool while preserving essential macromolecular information. More importantly, the method achieves substantially higher {\color{black}RMP} for MPF mapping, which is critical given the subtle signals from macromolecules in vivo. The effectiveness of the proposed method was validated through Bloch-McConnell simulations, as well as phantom and in vivo experiments. Overall, the proposed MPF-PSL approach addresses key limitations of spin-lock-based MPF mapping, enhancing its practicality and reliability for clinical applications.

\section*{Acknowledgments}

This study was supported by a grant from the Research Grants Council of the Hong Kong SAR (Project GRF 14213322), and a grant from the Innovation and Technology Commission of the Hong Kong SAR (Project No .MRP/046/20x).  
The research was conducted in part at CUHK DIIR MRI Facility, which is jointly funded by Kai Chong Tong, HKSAR Research Matching Grant Scheme and the Department of Imaging and Interventional Radiology, The Chinese University of Hong Kong.

The authors would like to thank Zijian Gao for preparing the phantom used for validation.

\section*{Conflict of interest}

The authors declare no potential conflict of interests.

\bibliography{MRM-AMA}

\section*{Supporting information}
The following supporting information is available as part of the online article:

\vskip\baselineskip\noindent
\textbf{Figure S1.}
{\color{black}Relationship between $R_\mathrm{mpfsl,pul}$ and the frequency offset (FO) $\Delta\omega^\mathrm{(1)}$, frequency of spin-lock (FSL) $\omega_1^\mathrm{(1)}$ and N, where N is a constant scaling factor such that $\Delta\omega^\mathrm{(2)}=N\Delta\omega^\mathrm{(1)}$ and $\omega_1^\mathrm{(2)}=N\omega_1^\mathrm{(1)}$. A, The $R_\mathrm{mpfsl,pul}$ signal as a function of FO, while fixing N = 4 and FSL = 50 Hz, 100 Hz, and 200 Hz, respectively.  B, The $R_\mathrm{mpfsl,pul}$ signal as a function of FSL, while fixing N = 4 and FO = 500 Hz, 1000 Hz, and 2000 Hz, respectively.  C, The $R_\mathrm{mpfsl,pul}$ signal as a function of N, while fixing FSL = 100 Hz and FO = 500 Hz, 1000 Hz, and 2000 Hz, respectively.  D, The $R_\mathrm{mpfsl,pul}$ signal as a function of FO and FSL, while fixing N = 4.}

\vskip\baselineskip\noindent
\textbf{Figure S2.}
{Off-resonance pulsed $R_\mathrm{1\rho}$ knee map for a volunteer. (A) Low RF pulse setting: $\omega_\mathrm{1} = 80 \cdot 2\pi \text{ rad}$ and $\Delta\omega = 800 \cdot 2\pi \text{ rad}$.  (B) High RF pulse setting: $\omega_\mathrm{1} = 350 \cdot 2\pi \text{ rad}$ and $\Delta\omega = 3500 \cdot 2\pi \text{ rad}$. ROIs FC-C, TC-A, TC-P correspond to the central regions of the femoral cartilage, and the anterior and posterior regions of the tibial cartilage, respectively; (C, D) The low- and high-RF cartilage signals from the ROIs were separately fitted to a mono-exponential model.}
\vskip\baselineskip\noindent
{\color{black}\textbf{Derivation S1.}
{T\lowercase{he expression of} $R_\mathrm{\lowercase{mt}}$}}

\vskip\baselineskip\noindent
\textbf{Derivation S2.}
{T\lowercase{ransient state relationship with $\MakeUppercase{M}_\mathrm{za}(t)$ and $\MakeUppercase{M}_\mathrm{zb}(t)$}

\vskip\baselineskip\noindent
\textbf{Derivation S3.}
{$R_\mathrm{\lowercase{mpfsl}}$ \lowercase{increases with} TSL}

\vskip\baselineskip\noindent
{\color{black}\textbf{Table S1.}
{$T_\mathrm{1\MakeLowercase{a}}$ \MakeLowercase{and} $T_\mathrm{2\MakeLowercase{a}}$ \MakeLowercase{values of the phantom with and without} M\MakeLowercase{n}C\MakeLowercase{l}$_2$}}

\nocite{*}

\end{document}